# Application of Spherical Convolutional Neural Networks to Image Reconstruction and Denoising in Nuclear Medicine

Amirreza Hashemi, Yuemeng Feng, Arman Rahmim, *Senior Member, IEEE,* and Hamid Sabet, *Member, IEEE*

*Abstract*— This work investigates use of equivariant neural networks as efficient and high-performance frameworks for image reconstruction and denoising in nuclear medicine. Our work aims to tackle limitations of conventional Convolutional Neural Networks (CNNs), which require significant training. We investigated equivariant networks, aiming to reduce CNN's dependency on specific training sets. Specifically, we implemented and evaluated equivariant spherical CNNs (SCNNs) for 2- and 3-dimensional medical imaging problems. Our results demonstrate superior quality and computational efficiency of SCNNs in both image reconstruction and denoising benchmark problems. Furthermore, we propose a novel approach to employ SCNNs as a complement to conventional image reconstruction tools, enhancing the outcomes while reducing reliance on the training set. Across all cases, we observed significant decrease in computational cost by leveraging the inherent inclusion of equivariant representatives while achieving the same or higher quality of image processing using SCNNs compared to CNNs. Additionally, we explore the potential of SCNNs for broader tomography applications, particularly those requiring rotationally variant representation.

## I. INTRODUCTION

Artificial intelligence (AI) has emerged as a promising technology in the medical imaging applications by enhancing diagnosis, treatment, and patient outcomes. It has shown immense potential to improve the quality of medical images in several ways such as enhancing spatial resolution, noise reduction, and lowering acquisition time [1], [2], [3]. Among AI solutions, convolutional neural networks (CNN) have played a prominent role as a powerful framework in medical imaging tasks. CNNs have been utilized in a wide range of research and application areas, spanning instrumentation, data acquisition, image reconstruction, and post-reconstruction enhancement [4], including image denoising while preserving details of medical images [5], [6].
In recent years, the use of CNN for reconstruction has extended to radiation-imaging modalities, such as PET, CT and SPECT, with significant opportunities and ongoing efforts in image quality and computational efficiency [6], [7], [8]. Recent progresses in deep learning CNN have focused on enhancing data fidelity and computational efficiency through learning-based models, serving as complements to conventional reconstruction algorithms like maximum likelihood expectation maximization (MLEM) [9], [10]. However, the enhancement of conventional CNNs has been hampered by many existing limitations that include overfitting, limited interpretability, limited ability to handle non-Euclidean spaces (e.g. image on the sphere) and missing or insufficient data, and being computationally expensive.

A common solution has been to employ augmented approaches that excel with large datasets, capable of extracting complex localizations of important features through interpolation and extrapolation. A well-known example of such approaches is the U-Net family [11], which utilizes an encoder-decoder architecture to extract essential features from the dataset. Although these approaches demonstrate success in handling complex and large datasets, they are less efficient when applied to limited and small datasets. U-Net models tend to overfit when trained on small datasets due to their large number of parameters, resulting in poor generalization to new data [12].

Furthermore, it is important to recognize that symmetry plays a crucial role in many anatomical structures and objects, and is clearly evident in medical images.[13], [14]. These symmetrical representations can be leveraged in the creation of training set representation of neural network model. In this regard, recent studies [15],[16], [17] on AI-based image reconstruction of PET and SPECT have suggested that the inclusion of the symmetrical rotation of training data can enhance the image quality and lower the computational cost. However, the improvements of the recent studies were limited to stacked representation of a single rotation of the training set. Additionally, it has been shown that the rotational invariance of the denoising filters leads to promising outcomes, yielding high-quality image outputs [18]. Here, our objective is to extend upon the recent progress and show the advantage of

A. Hashemi, Y. Feng, and H. Sabet are with the Division of Nuclear Medicine & Molecular Imaging, Department of Radiology, Massachusetts General Hospital & Harvard Medical School, Boston, MA, USA (email: sahashemi, yfeng16, hsabet@mgh.harvard.edu). A. Rahmim is with the Departments of Radiology and Physics, University of British Columbia, Vancouver, BC V6T 1Z4, Canada (email:arman.rahmim@ubc.ca) This work is partially supported by NIH Grants No. R01HL145160 & R01EB034785.

equivariance in machine learning-based medical imaging processing. In machine learning of symmetrical datasets, behavioral properties exhibit distinct transformation characteristics when subjected to translation, reflection, and rotation with respect to small local sample representatives.

In the present work, we investigate the use of equivariant spherical CNN (SCNN) for medical imaging applications and particularly for problems where the representative domain is symmetrical (or spherical) such as brain images. We show variational invariances in training set play a key role on the performance of CNNs in medical imaging applications, and therefore, approaches such as SCNN are well suited because the network model is adapted to rotational, translational and permutational invariances. To prove our proposition, we show the efficiency of SCNNs for denoising and reconstruction for several problems in 2- and 3- dimension (2- and 3-D). Additionally, we utilize the SCNN as an alternative method to conventional image reconstruction to improve the performance and quality of the overall outcome based on a limited dataset. Ultimately, we show that SCNNs and equivariants network are a viable option that reduce performance dependence on the training set for CNN-based medical imaging algorithms.

## II. METHOD

SCNNs were first introduced [19], [20], [21], [22] as specific subsets of CNNs that efficiently encodes the symmetries of data by leveraging the properties of orthogonal group representations on spherical harmonics. In this regard, equivariant SCNN utilizes spherical harmonics and convolutional operations to extract features from the spherical signals. Spherical harmonics capture the harmonic components of the input signal, while convolutional operations perform localized computations on the spherical surface. In addition, the sphere has a constant positive curvature, which means that the distance between any two points on the sphere is always less than the distance between the same two points on a flat surface. This property allows SCNNs to be more efficient at processing data than conventional CNNs, which are designed for flat, Euclidean spaces.

To achieve equivariance, SCNN applies a group-equivariant convolutional layer which operates on feature vector fields associated to the actions of $(\mathbb{R}^n, +) \rtimes G$ where $G$ is the group transformation function and $(\mathbb{R}^n, +)$ is the space of semidirect product of translations. The orthogonal group representation $\rho: G \to \mathbb{R}^{d \times d}$ is the geometric feature field where $d$ is the feature vector dimension and feature fields are a transformation map $f: \mathbb{R}^n \to \mathbb{R}^d$ that transforms in space of $(\mathbb{R}^n, +) \rtimes G$. According to [23], the $\rho$-field transforms under the induced representations is given as

$$\left(\left[\text{Ind}_G^{(\mathbb{R}^n,+) \rtimes G} \rho\right](tg) \cdot f\right)(x) := \rho(g) \cdot f(g^{-1}(x-t)) \quad (1)$$

where it transforms the feature fields $f$ by moving feature vectors spatially from $g^{-1}(x-t)$ to $x$ and is based on the $\rho(g)$. Vector fields take various forms, for example when transformation is based on the consistent scalar value, $\rho(g) = 1$, feature vector correspond to trivial representation and when the transformation in under the action of $g$, $\rho(g) = g$, feature vector correspond to standard or regular representation.

In the most general form, the full feature space transforms are convolutions with G-steerable kernels using the full feature space transforms where the linear G-steerability constraint is defined as

$$K(g \cdot x) = \rho_{\text{out}}(g) K(x) \rho_{\text{in}}(g)^{-1} \quad (2)$$

for $\forall g \in G, x \in \mathbb{R}^n$ and $K: G \to \mathbb{R}^{d_{\text{out}} \times d_{\text{in}}}$, and where $\rho_{\text{in}}: G \to \mathbb{R}^{d_{\text{in}} \times d_{\text{in}}}$ and $\rho_{\text{out}}: G \to \mathbb{R}^{d_{\text{out}} \times d_{\text{out}}}$ are the input and output fields respectively. In the SCNNs, a basis of the vector space of G-steerable kernels is found to parameterize conventional Euclidean convolutions. This parametrization satisfies the steerability constraints which in the vectorized form is simplified as

$$k(g \cdot x) = [(\rho_{\text{in}} \otimes \rho_{\text{out}})(g)] k(x), \quad \forall g \in G, x \in X \quad (3)$$

where $(\rho_{\text{in}} \otimes \rho_{\text{out}})(g)$ is the Kronecker product of two matrices of input and output feature vector fields and $X$ is a general space. The steerable basis is square-integrable function on $X$ given as $L^2(X)$ and is a collection of orthogonal functions denoted by the property of $Y_{ji}(g \cdot x) = \rho_j(g) Y_{ji}(x)$ where $Y_{ji}$ is the stack of the orthogonal basis. Hence the compact form of the vectorized G-steerable kernel function is calculated to be as following:

$$K_{jisr}(x) = \left[\text{CG}_s^{j(l^\dagger J)}\right]^\dagger \cdot c_r^j \cdot Y_{ji}(x) \quad (4)$$

where dots are the convolution matrix operations and $\left[\text{CG}_s^{j(l^\dagger J)}\right]^\dagger$ are the conjugate gradient of Clebsch-Gordan coefficients, and $c_r^j$ are the basis coefficients in matrix format. The proof and details of the kernel function are discussed in previous works [23], [24].

The kernel functions represent the spherical harmonics that satisfy the G-steerable constraints and results in equivariant convolution kernels. The requirements for the SCNN are input and output feature vectors, $\rho_{\text{in}}$ and $\rho_{\text{out}}$, and steerable basis $Y_{ji}$. The implementation of equivariant SCNN is to i) compute the basis of steerable kernels, $c_r^j$ for endomorphism space of all $\rho_j$'s ii) decompose the steerable kernel based on the basis with learned expansion coefficients iii) operate the convolution on subgroup steerable space. The typical SCNN decomposes basis using the independent feature fields and operates the convolution under steerable constraints shown in equations (3) and (4).

### A. Network Implementations: SCNN and CNN in 2-D and SCNN in 3-D

Our 2-D SCNN is built of three layers that include input, hidden, and output layers with 8 rotational actions on the group space. Each layer is connected to the inner batch normalization and rectified linear unit (ReLU) activation. Input and hidden layers have trivial and regular representative functions respectively with a kernel size of 3 and the output layer has an irreducible representation function with kernel size of one (for the details of SCNN descriptions and more information on the representative functions, see [23]). For the sake of comparison, a 2-D deep learning conventional CNN was also implemented with 5 layers (3 hidden layers), 64 channels, and ReLU activation [25].

Our 3-D SCNN is built of three layers that include input, hidden, and output layers with theoretically unlimited rotational actions on the group space. Each layer is connected to the generalized 3-D batch normalization function followed by norm nonlinearity activation function. Input and hidden layers have trivial and spectral-regular representative functions with a kernel size of 3. And the output layer has an irreducible representation function with one kernel. Spectral regular representative is when the input and output fields exude Fourier coefficients. Also, the loss function is defined as mean squared error. More details on the mathematical meanings of representatives are presented in [23], [24]. Figure 1 presents a schematic overview of the 2-D and 3-D networks for an example input image. Fundamentally, the algorithm operates similarly in both 2-D and 3-D networks, focusing on the local environments of the inputs. The primary distinction lies in the 3-D network's use of spectral regular representatives in the hidden layers, as opposed to regular representatives in the 2-D network.

We utilized the PyTorch framework [26] and the escnn package, with detailed elaborated elsewhere [23], [24]. Denoising and reconstruction simulations were performed on Linux Ubuntu v20.04 with an Intel Xeon E5-2687W 3.1GHz CPU, 128GB RAM, and an NVIDIA TITAN RTX GPU card with 24GB of memory.

For the 2-D single-dataset case, we utilized a brain image from the Scikit-learn brain image library. To evaluate image reconstruction performance, the system matrix transformed the truncated brain image with 64×64 pixels into a sinogram of the same size using the Radon transform (as shown in Figure 4a). The sinogram was then transformed into the input feature function (trivial representation, $\rho_j(g) = 1$) via a geometric tensor mapping function of PyTorch before being fed into the SCNN network. The forward propagation model is integrated into the loss function of the neural networks.

In 3-D case, to evaluate the performance of the SCNN on large datasets, we trained 3-D SCNN on a limited rotationally invariant dataset and test the model on a rotationally variant testing dataset. We utilize whole-brain PET data [27], which inherently includes rotationally variant 2-D images. The training is conducted on a small set of rotationally invariant images, while the testing is performed on a set of rotationally variant images. The training set comprises 10 randomly selected slices of rotationally invariant 2-D images, each with a pixel size of 110×110.

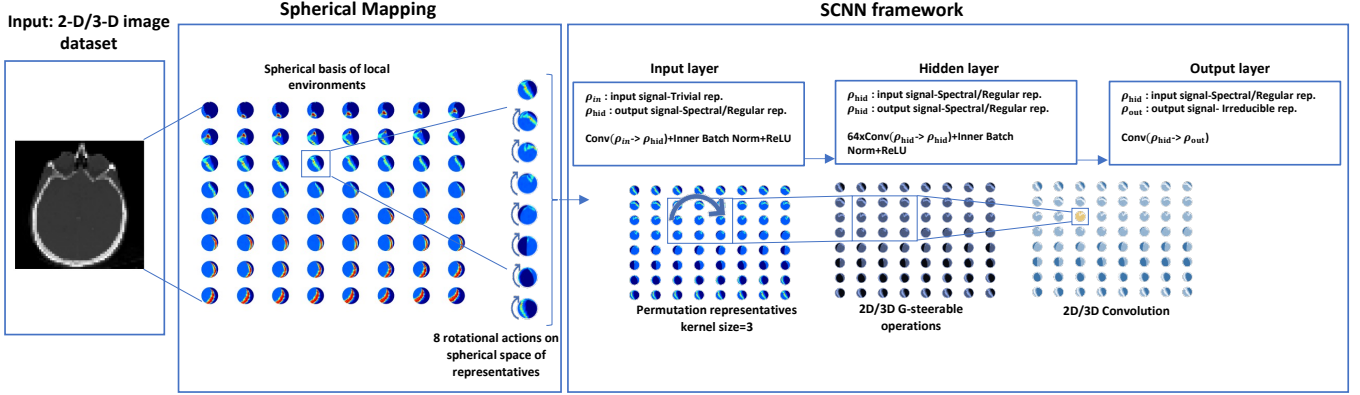

Fig. 1. Schematic view of our 2-D and 3-D SCNN implementation. The input image is mapped to the spherical spaces of its local representatives, with each spherical representative encompassing 8 rotational variations in the mapping space. The input, hidden, and output layers illustrate convolutional operations applied to linear representatives under steerable constraints, culminating in the irreducible representation of all equivariant representatives.

### B. Machine Learning Assisted Image Reconstruction

#### B.1. Monte Carlo simulation setup

PET simulations were conducted using GATE [28] involving a designed phantom comprising six regions of hot rods. These rods have a length of 5 mm and diameters measuring 6.5 mm, 5.5 mm, 4.5 mm, 3.5 mm, 2.2 mm, and 1.6 mm. Additionally, there is a central cold region with a diameter of 3 cm. The ratio between the hot rods and the warm background is set to 10, with the activity of the hot rods at 53kBq/cc and background at 5300 Bq/cc. The phantom was positioned in air, and a 10-minute acquisition was performed. The simulated whole-body PET geometry consists of 5 rings and each ring has 44 detector blocks. Each detector block consists of 2.1 x 2.1 x 20 mm³ pixels with 2.2 mm pixel pitch, two depth of interaction (DOI) levels. The reconstructed volume has a voxel size of 0.8 x 0.8 x 0.8 mm³. Target coincidence time resolutions (CTRs) is tested for the system with 50 ps FWHM.

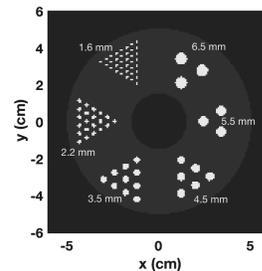

Fig. 2. The designed phantom utilized in this work consists of 6 regions of hot rods.

### B.2. Machine learning framework

In this study, we implemented and investigated SCNN-assistance for image reconstruction. The algorithm consisted of two main steps. First, an early list mode MLEM reconstruction (using only second iteration) was performed based on the simulation data. Then, a subsequent SCNN reconstruction was carried out using the 2-D central slice of the MLEM reconstruction as the input to the SCNN. The PET system matrix in list mode MLEM was calculated on the fly [29], taking into consideration the probability of detecting one coincidence event emitted from a voxel by the line of response connecting the opposing detector pixels.

### III. RESULTS

#### A. Comparison between SCNN and CNN

In this section, we provide comparison of our proposed SCNN with conventional CNN for denoising and reconstruction problem of a single dataset. We s use the self-supervised deep learning followed by [22] to introduce the advantages and performance of SCNN over the conventional CNN.

#### A.1. Denoising Example for 2-D Single Dataset

To evaluate the performance of our SCNN model, first we compare its results with those of conventional CNNs for denoising a noisy brain image. Figure 3a presents the true image, while Figure 3b displays the corresponding noisy image with Poisson noise. Both CNN and SCNN models are trained for 2000 epochs, and the denoised images are shown in Figure 3c and Figure 3d at the 2000th epoch. Additionally, the denoised images at the 500th epoch are illustrated in Figure 3e and Figure 3f for CNN and

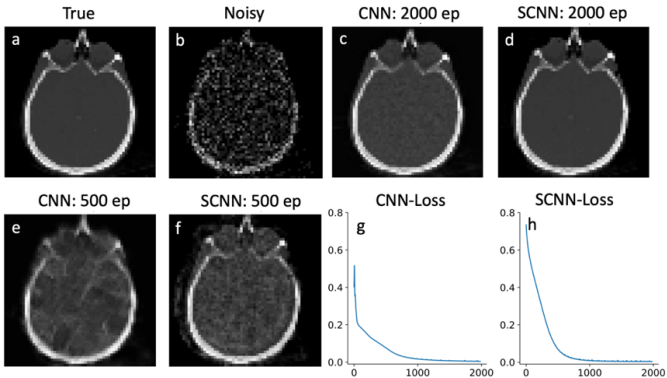

Fig. 3. Denoising results for brain image. Plots a and b show true and noisy images, comparison between CNN and SCNN are shown in plots c-f for 2000 and 500 epochs. The loss functions for CNN and SCNN are shown in plots g and h respectively.

SCNN, respectively. The loss function, measured by the mean square error (L2 norm), is employed to assess the training progress. The loss function versus epoch plots for CNN and SCNN are presented in Figure 3g and Figure 3h, respectively. We also compared the results in terms of quality metrics of structural similarity (SSIM) and mean square error (MSE) in Table 1 to draw a quantifiable distinction between the results. Regarding the computational aspect, both CNN and SCNN models exhibit similar computational times for 2000 epoch iterations when executed on the GPU, measuring 104 seconds for CNN and 106 seconds for SCNN in the case of denoising.

Table 1: Comparison SSIM and MSE for 2-D denoising

| epochs | SSIM | | MSE | |
|---|---|---|---|---|
| | CNN | SCNN | CNN | SCNN |
| 500 | 0.59 | 0.71 | 0.0907 | 0.019 |
| 2000 | 0.82 | 0.96 | 0.0041 | 0.00053 |

#### A.2. Image Reconstruction Example for 2-D Single Dataset

To evaluate the performance for image reconstruction, the sinogram of the benchmark brain image is generated using the Radon transform which is shown in Figure 4a, and the forward propagation model is integrated into the loss function of the neural networks. The reconstruction results for CNN and SCNN are presented for two epochs at 5000 and 500 iterations, while the corresponding loss function versus epoch plots are displayed in Figures 4d and 4g. The minimum values of the loss function for SCNN and CNN converge to 0.015 and 0.14, respectively. We also showed the quality metric comparison between CNN and SCNN in Table 2 at three epoch iterations of 500th, 2000th, and 5000th. In terms of computational time, CNN takes 766 seconds, while SCNN requires 784 seconds to complete 5000 epochs. Additionally, the computational time for SCNN at 500 epochs is estimated to be 87 seconds.

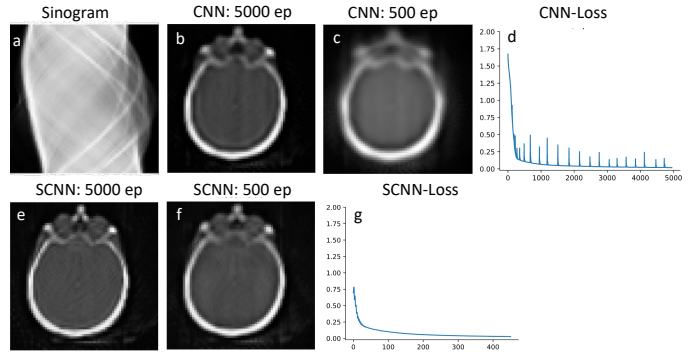

Fig. 4. Reconstruction results using sinogram input that is shown in plot a. Comparison between the reconstructed image at 5000 and 500 epochs are shown in plots b and c for CNN and plots e and f for SCNN. Plots d and g are loss function vs epoch for CNN and SCNN respectively.

Table 2: Comparison SSIM and MSE for 2-D reconstruction

| epochs | SSIM | | MSE | |
|---|---|---|---|---|
| | CNN | SCNN | CNN | SCNN |
| 500 | 0.59 | 0.73 | 0.0797 | 0.02 |
| 2000 | 0.82 | 0.97 | 0.0042 | 0.001 |
| 5000 | 0.96 | 0.97 | 0.00097 | 0.00048 |

#### B. SCNN for Large Dataset

In this section, we examine the efficacy of the SCNN for datasets of the whole brain PET data with rotational variance [30] and extension to 3-D SCNN and comparing with the previous study [5].

*B.1. Denoising Example for Rotationally Variant Dataset*

Figure 5 illustrates three different images, showcasing both noisy and noiseless (true) images, along with the loss function for the training set. The training set exhibits a consistent and monotonically decreasing loss function, similar to that observed in single data denoising and reconstruction tasks. Additionally, we present the results for three different testing examples where the images are rotationally variant compared to the training images. The loss function in these cases shows noticeable oscillations during the early epochs but eventually converges to a monotonically decreasing behavior in the subsequent epochs.

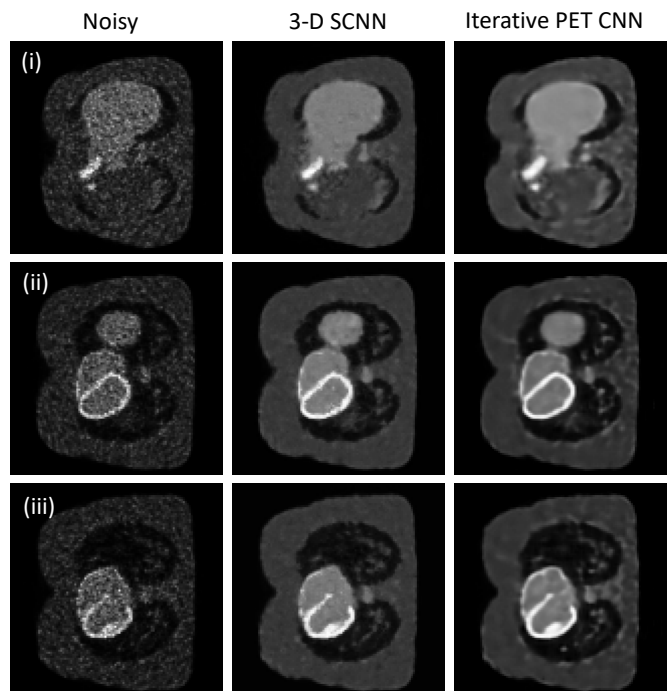

Fig. 6. Comparison of 3-D SCNN for three selected slices (corresponding slices are placed in rows of i, ii, and iii) of the input noisy images, 3-D SCNN outputs and the outputs of alternative iterative PET CNN reconstruction at 100th iteration [5].

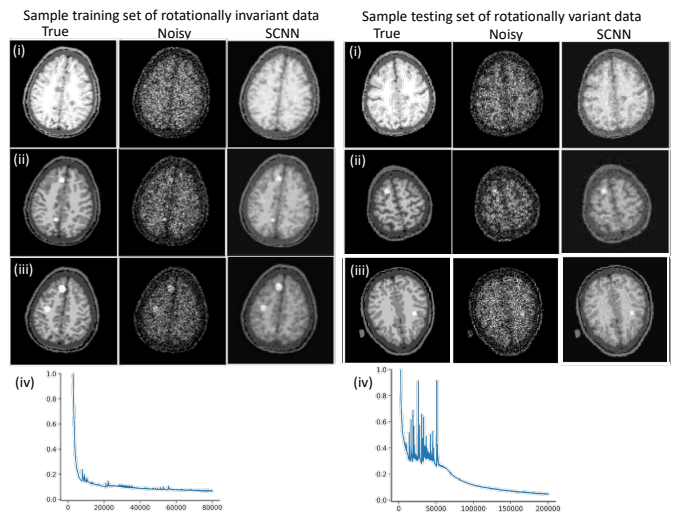

Fig. 5. Denoising results based on the limited rotationally invariant and testing on rotationally variant dataset. The results in (i) to (iii) show three examples of the training and testing datasets and (iv) shows the loss function vs epochs for the corresponding dataset. The MSE-SSIM of training samples are 0.064-0.89, 0.067-0.89, and 0.069-0.84 and MSE-SSIM of testing samples are 0.073-0.78, 0.77-0.89, and 0.071-0.83.

*B.2. Denoising Example for 3-D Data*

For 3-D case, we examine our network with the simulation study of XCAT phantom[31]. This approach builds on previous work[5] that incorporated neural network models into iterative PET image reconstruction frameworks. Specifically, we embed our network within an iterative reconstruction process, similar to these prior studies, to assess its performance and capabilities in handling 3-D imaging scenarios. The dataset includes image matrix of the size 128×128×49 and the voxel size of 3.27×3.27×3.27 mm$^3$, with the Poisson noise level that was generated to replicate the real-world dataset. The noisy results served as inputs for our denoising 3-D SCNN model. To facilitate comparison, we selected and presented three slices of the noisy inputs in Figure 6. Our 3-D SCNN model was exclusively initialized with the noisy dataset, and the denoising problem were run for 60,000 epochs until the loss value approached near zero. Additionally, we displayed the corresponding outputs of the image reconstruction obtained from iterative PET reconstruction, which was executed for 100 iterations and can be found in [32] showcase an improved version of the CNN. Therefore, we demonstrate the effectiveness of the SCNN in the augmented model when applied to a large 3-D database.

*C. SCNN-Assisted Efficient Image Reconstruction Based on Small Dataset*

Figure 7a and 7b display the outcomes of SCNN-assisted reconstruction, where the reconstruction process begins with the initial two iterations of MLEM-based reconstruction which is demonstrated in Figure 7e. Only a single central slice of the MLEM-reconstructed image and its sinogram is imported as a training set for the SCNN reconstruction algorithm. A similar framework is employed for CNN, and the results are presented in Figure 7c and 7d. The SCNN-assisted reconstruction demonstrates high accuracy in reconstructing rods with diameters ranging from 3.5 mm to 6.5 mm. However, the reconstruction appears blurry for the rods with diameters of 2.2 mm and 1.6 mm. It is worth noting that the SCNN loss function exhibits a consistent and steady decrease with minimal instability up to the point where the minimum loss value is achieved (Figure 7b).

IV. DISCUSSION

The results obtained from SCNN demonstrate higher accuracy in both image denoising and reconstruction problems compared to conventional CNNs. This improvement can be attributed to the utilization of an equivariant dataset as input for SCNN, which is not present in conventional CNNs. To illustrate this point, Figure 8 showcases eight equivariant rotational representative actions applied to denoising cases throughout a single epoch iteration of SCNN. Consequently, each training input undergoes transformation within the rotational space of these representative actions. This inclusion of equivariant representative actions reduces SCNN's reliance on extensive data augmentation to achieve robustness to transformations, thereby enhancing output accuracy. Unlike traditional CNN-based approaches, which often rely on data augmentation by adding rotated and translated versions of input datasets to the training set [5], [7], [33] SCNN operates directly on the spherical space. This allows the convolution to inherently account for rotational and permutational invariance within the architecture. By embedding these symmetries at a structural level, SCNN ensures that the model generalizes effectively to transformations without requiring explicit exposure to all variations during training. This approach not only enhances invariance and efficiency but also mitigates the limitations of manual augmentation, which can be subjective, computationally expensive, and dependent on the developer's experience.

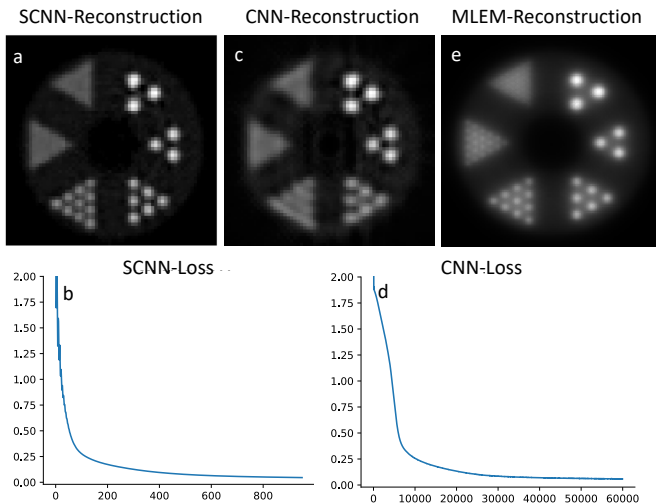

Fig. 7. Results of SCNN assisted reconstruction is shown in plots a and, loss result of the first 1000 epochs of the loss result are shown in plot b. Results of CNN assisted reconstruction is shown in plots c and, loss result is shown in plot d. Plot e shows the MLEM reconstruction at second iteration.

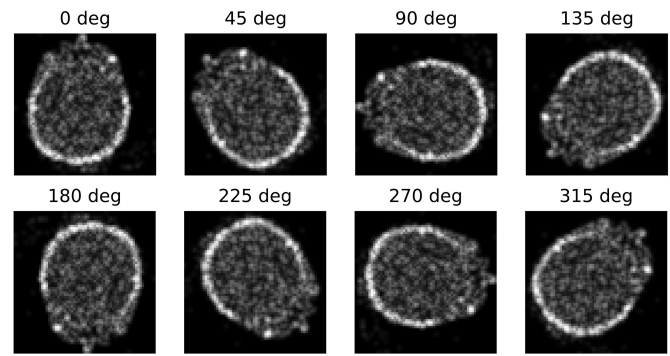

Fig. 8. Rotational representative actions show output transforms of a noisy brain image.

In addition, SCNN exhibits faster convergence, reaching a near-zero value for the loss function in fewer iterations. Specifically, in the 2-D denoising case, SCNN achieves a loss value equivalent to that of CNN at 2000 epochs by only 1000 epochs. For the 2-D reconstruction case, SCNN's loss value converges to a lower value of 0.02 at 500 epochs, whereas CNN requires 5000 epochs to converge to a higher value of 0.14. Figures 4b and 4f provide a visual comparison, demonstrating that SCNN achieves higher quality image reconstruction ten times faster in terms of epoch count and nearly nine times faster in terms of computational time. Additionally, the 2-D SCNN loss function exhibits a consistent and monotonous decrease, with minimal instability, until it reaches the minimum loss value. This smooth and stable behavior is a key factor contributing to the rapid convergence of the SCNN model. In contrast, when considering the case of 2-D CNN, the loss function displays some instabilities during the early epochs. These instabilities can be attributed to the primary limitations of conventional CNN models in effectively handling non-Euclidean spaces, as evident in the denoising of noisy images.

We demonstrated that the SCNN can be effectively trained on a limited set of rotationally invariant data and then applied to a rotationally variant dataset. However, the SCNN's consistent behavior of a monotonically decreasing loss function is observed in the early training epochs on the rotationally variant dataset. We observed that this behavior is preserved across testing as well. During the testing phase, the results obtained after a number of epochs on the sample data demonstrates comparable performance, and in the case of larger lesions, it exhibits superior quality and recoverability compared to the results obtained from iterative PET CNN. While the image contrast is marginally reduced in the 3-D SCNN results, it effectively preserves intricate edge details in the reconstructed images. Notably, the denoised SCNN sliced images exhibit reduced blurriness, particularly in the connecting regions of the contrasting lesions. We should note that the 3-D SCNN loss function follow similar pattern as the 2-D SCNN, however larger number of epochs was used to converge these results to compare with the 2-D SCNN. This is partially related to more intricate input dataset. Also 3-D SCNN is a newly developed package with a suboptimal simulation performance, and the limited number of representative functions currently available influences the computational performance at this stage.

Regarding assisted image reconstruction, SCNN-assisted approach shows high accuracy in reconstruction of rods with diameters ranging from 3.5 mm to 6.5 mm. However, the reconstruction is blurry for the rods with diameters of 2.2 mm and 1.6 mm. It is noticeable that the SCNN loss function consistently and steadily decreases with minimal instability until it reaches the minimum loss value, as depicted in Figure 7b. In contrast, the CNN-assisted results showcase accurate reconstruction of rods with diameters of 5.5 mm and 6.5 mm, while the 4.5 mm diameter rods are not well resolved, and the remaining rods are not well-resolved. The convergence rate of the CNN loss is significantly slower when compared to the SCNN loss. It is worth mentioning that increasing the number of epochs did not lead to improvements in CNN results. In terms of computational cost, the SCNN and CNN-assisted reconstructions take 870 seconds and 821 seconds, respectively, to complete 60,000 epochs. Notably, SCNN achieves convergence in just 1,000 epochs with a computational time of 15 seconds. Therefore, the computational cost is significantly reduced when compared to list mode MLEM, where each iteration takes approximately 2,112 seconds to compute on one CPU thread.

Overall, SCNN offers a simple and straightforward architecture with few adjusting parameters that perform more efficiently than conventional CNNs. We demonstrated that the SCNN is particularly efficient for the limited dataset such as self-supervised deep learning based on a single dataset, or the limited rotationally invariant dataset. While previous works have focused on large dataset [5], [6], [11], [30]. We focused our comparison on CNN and SCNN to clarify the differences between the conventional convolution operator and the spherical convolution operator. In addition to the previously mentioned applications of SCNN, which primarily demonstrated its relevance to limited datasets, SCNN can also be combined with the encoder-decoder approach to leverage spherical convolution in large datasets which remains possible avenue of future research.

## V. Conclusion

In this work, we demonstrated the effectiveness of equivariant SCNNs in improving image quality for various examples such as denoising and image reconstruction of 2-D CT brain image, while significantly reducing computational costs compared to conventional CNNs. We also extend our model to 3-D and examined the performance of newly developed 3-D SCNN on 3-D noisy PET dataset. Additionally, our study showcased the successful combination of equivariant SCNN with image reconstruction techniques such as list mode MLEM reconstruction, achieving high accuracy and computational efficiency even when utilizing a small subset of data in SCNN reconstruction. These findings highlight the potential of SCNN for broader medical imaging applications, especially in scenarios where input data may be limited and involve a spherical field of view (FOV) [34], [35], [36], [37] and particularly in modalities with omnidirectional outputs like brain PET or cardiac dedicated scanners. Hence, further investigations are planned to explore the applicability of SCNNs for such problems in the future.